\documentclass[prb,twocolumn]{revtex4}   % Physical Review B
\usepackage{graphicx}

\begin{document}

%%%%%%%%%%%%%%%%%%%%%%%%%%%%%%%%%%%%%%%%%%%%%%%%%%%%%%%%%%%%%%%%%%%%%%%
\title
{
Stability of a metallic state in the two-orbital Hubbard model
}
%%%%%%%%%%%%%%%%%%%%%%%%%%%%%%%%%%%%%%%%%%%%%%%%%%%%%%%%%%%%%%%%%%%%%%%

\author
{   							
Akihisa Koga, Yoshiki Imai and Norio Kawakami
}

\affiliation
{
Department of Applied Physics, Osaka University, Suita, Osaka 565-0871, Japan
}

\date{\today}

\begin{abstract}
Electron correlations in the two-orbital Hubbard model at 
half-filling are investigated by combining dynamical mean field
theory with the exact diagonalization method. We systematically study
how the interplay of the intra- and inter-band Coulomb interactions, 
together with the Hund  coupling, affects the metal-insulator transition.
It is found that if the intra- and inter-band Coulomb interactions
are nearly equal, the  Fermi-liquid state is 
stabilized due to orbital fluctuations up to fairly large 
interactions, while the system is immediately driven to the Mott
insulating phase away from this condition. 
The effects of the isotropic and anisotropic  Hund coupling  are also
addressed.
\end{abstract}

\pacs{Valid PACS appear here}% 

\maketitle

\section{Introduction}
Strongly correlated electron systems with multi-orbital
bands  have attracted current interest. 
Typical examples are the manganite $\rm La_{1-x}Sr_xMnO_3$\cite{Tokura} and 
the ruthenate $\rm Sr_2RuO_4$,\cite{Maeno}
where striking phenomena such as the colossal magnetoregistance and the 
triplet pairing superconductivity were observed, 
stimulating intensive 
theoretical and experimental 
investigations.\cite{Tokura2,RiceSigrist}
Common physics in the above compounds is that 
orbital degrees of freedom together with the Hund coupling
 play a key role in realizing novel phenomena at low 
temperatures.

Another interesting example is the vanadium oxide $\rm LiV_2O_4$,
where the heavy fermion behavior was observed 
at low temperatures. \cite{Kondo}
It has been suggested that the large mass enhancement
 in this system may originate from
geometrical frustration.\cite{Kaps,Isoda,Fujimoto}
More recently, it has been pointed out that degenerate 
orbitals are also important to understand the heavy fermion behavior in 
$\rm LiV_2O_4$.\cite{Tsunetsugu,Yamashita}

It is thus desirable to discuss how the metallic ground state is 
affected by degenerate orbitals and also by correlations among  them.
Although the effects of the orbital degeneracy have been explored  extensively,
\cite{Khaliullin,Ishihara,Motome,Hasegawa,Fresard,Klenjnberg,
Takimoto,Bunemann,Momoi,Rozenberg,Held,Han,QMC,Kotliar}
the stability of a metallic state due to orbital fluctuations
has not been studied systematically.
In this paper, we study the two-orbital Hubbard model at half-filling 
to discuss how the interplay of the intra-band interaction,
the inter-band interaction and the Hund coupling affects the 
stability of a metallic phase against an insulating phase.
In particular, by using dynamical mean-field theory (DMFT),
we show  that orbital fluctuations stabilize a heavy fermion state
 when the intra- and inter-band
Coulomb interactions are nearly equal.

This paper is organized as follows.  
We introduce the model Hamiltonian and briefly summarize 
the formulation based on DMFT in Sec. \ref{sec2}. 
 In Sec. \ref{sec3}, we then discuss the metal-insulator transition  
by  combining  DMFT with the exact diagonalization 
method, and  clarify how the 
intra- and inter-band couplings  affect the stability of 
the Fermi liquid phase. 
%% We demonstrate the remarkable role played by orbital fluctuations 
%%to stabilize heavy Fermi liquid
%%in the vicinity of the metal-insulator transition.
The effects of the  Hund coupling are addressed in Sec. \ref{sec4}. 
A brief summary is given in Sec. \ref{Summary}.

%%%%%%%%%%%%%%%%%%%%%%%%%%%%%%%%%%%%%%%%%%%%%%%%%%%%%%%%%%%%%%%%%%%%%%%%
\section{Model and Method }\label{sec2}
%%%%%%%%%%%%%%%%%%%%%%%%%%%%%%%%%%%%%%%%%%%%%%%%%%%%%%%%%%%%%%%%%%%%%%%%

\subsection{Two-orbital Hubbard Hamiltonian}
%%%%%%%%%%%%%%%%%%%%%%%%%%%%%%

We consider a correlated electron system with two-fold degenerate orbitals,
which is described by the following Hubbard Hamiltonian,
%%%%%%%%%%%%%%%%%%%%%%%%%%
\begin{eqnarray}
H&=&\sum_{\stackrel{<i,j>}{\alpha,\sigma}}
t_{ij} c_{i\alpha\sigma}^\dag c_{i\alpha\sigma}
+U\sum_{i\alpha}n_{i\alpha\uparrow}n_{i\alpha\downarrow}\nonumber\\
&+&U'\sum_{i,\sigma,\sigma'}n_{i1\sigma}n_{i2\sigma'}
+J\sum_{i}{\bf S}_{i1}\cdot{\bf S}_{i2},
\end{eqnarray}
%%%%%%%%%%%%%%%%%%%%%%%%
where $c_{i\alpha\sigma}^\dag (c_{i\alpha\sigma})$ 
creates (annihilates) an electron 
with  spin $\sigma(=\uparrow, \downarrow)$ and orbital
index $\alpha(=1, 2)$ at the $i$th site, 
and $n_{i\alpha\sigma}=c_{i\alpha\sigma}^\dag c_{i\alpha\sigma}$.
The corresponding spin operator is defined by 
${\bf S}_{j\alpha}=\frac{1}{2}\sum_{\sigma\sigma'}c_{j\alpha\sigma}^\dag 
{\bf \tau}_{\sigma\sigma'}
c_{j\alpha\sigma'}$, where ${\bf \tau}$ is the Pauli matrix.
$t_{ij}$ represents the transfer integral,
$U$ and $U'$ the intra-band and inter-band Coulomb interaction,
and $J$ the Hund coupling.

In the following, we will particularly focus on how
the interplay of $U$, $U'$ and $J$ affects the metal-insulator 
transition.

%%%%%%%%%%%%%%%%%%%%%%%%%%%%%%%%%%%%%%%%%%%%%%%%%%%%%%%%%%%%%%%%%%%%%%%%%%%%%%
\subsection{Dynamical mean-field theory}
%%%%%%%%%%%%%%%%%%%%%%%%%%%%%%%%%%%%%%%%%%%%%%%%%%%%%%%%%%%%%%%%%%%%%%%%%%%%%%

We make use of DMFT, \cite{Metzner,Muller,Georges,Pruschke}
which has been extensively used for the 
single-band Hubbard model,
\cite{Caffarel,Sakai,single1,single2,single3,single4,BullaNRG} 
the  two-band model,\cite{Caffarel,2band1,2band2,2band3,OnoED}
the periodic Anderson model,\cite{PAM,Mutou,Saso} etc.
This method has also been applied to the degenerate Hubbard model
by combining it with  the exact diagonalization,\cite{Momoi} 
 quantum Monte Carlo simulation,\cite{Rozenberg,Held,Han,QMC}
 and  iterative perturbation theory.\cite{Kotliar} 
 
In DMFT, the lattice model is mapped to
 an effective impurity  model, 
where local electron correlations are taken into account precisely. 
The lattice Green function is then obtained via self-consistent
conditions imposed on the impurity problem.
The treatment is exact in  $d\rightarrow\infty$ dimensions, and
even in three dimensions,  DMFT has successfully explained 
interesting physics such as the Mott metal-insulator transition.
We introduce here the impurity 
 Anderson Hamiltonian with two orbitals,
%%%%%%%%%%%%%%%%%%%%%%%%%%%%%%%%%%%%%%%%%%%%%%%%%%%%%%%%%%%%%
\begin{eqnarray}
H_{\rm imp}&=&\sum_{k,\alpha,\sigma} 
\epsilon_{k} c_{k\alpha\sigma}^\dag c_{k\alpha\sigma}
+\sum_{\alpha\sigma} E_f f_{\alpha\sigma}^\dag f_{\alpha\sigma}\nonumber\\
&+&\sum_{k,\alpha,\sigma}V_{k} 
\left(
c_{k\alpha\sigma}^\dag f_{\alpha\sigma}
+f_{\alpha\sigma}^\dag c_{k\alpha\sigma}
\right)\nonumber\\
&+&U\sum_\alpha f_{\alpha\uparrow}^\dag f_{\alpha\uparrow}
f_{\alpha\downarrow}^\dag f_{\alpha\downarrow}
+U'\sum_{\sigma\sigma'} f_{1\sigma}^\dag f_{1\sigma}
f_{2\sigma'}^\dag f_{2\sigma'}\nonumber\\
&+&J{\bf S}_{1}\cdot{\bf S}_{2},\label{eq:And}
\end{eqnarray}    
%%%%%%%%%%%%%%%%%%%%%%%%%%%%%%%%%%%%%%%%%%%%%%%%%%%%%%%%%%%%%
where $f_{\alpha\sigma}^\dag (f_{\alpha\sigma})$ 
 creates (annihilates) an electron with spin $\sigma$ 
in the $\alpha$-th orbital at the impurity site, and 
${\bf S}_{\alpha}=\frac{1}{2}\sum_{\sigma\sigma'}f_{\alpha\sigma}^\dag
{\bf \tau}_{\sigma\sigma'}f_{\alpha\sigma'}$.
Note that the effective parameters in the impurity model, 
such as the spectrum of host electrons $\epsilon_k$ and
the hybridization $V_k$, should be determined 
self-consistently so that the obtained results 
properly reproduce the original lattice problem. This
 will be done explicitly below.
We focus on the symmetric case with half-filled
bands by setting  $E_f=-\frac{1}{2}U-U'$, for simplicity. 

To discuss the stability of the Fermi liquid state 
in a normal metallic phase,
we introduce the Green function with two components 
${\bf \Psi}^\dag=\left(f_1^\dag \;\; f_2^\dag\right)$ as
%%%%%%%%%%%%%%%%%%%%%%%%%%%%%%%%%%%%%%%%%%%%%%%%%
\begin{eqnarray}
{\bf G}\left(\tau-\tau'\right)&=&-\langle g|T{\bf \Psi}\left(\tau\right) 
{\bf \Psi}^\dag\left(\tau'\right)|g \rangle,
\end{eqnarray} 
%%%%%%%%%%%%%%%%%%%%%%%%%%%%%%%%%%%%%%%%%%%%%%%
where $T$ is the time-ordering operator and $|g \rangle$ 
represents the ground state.
In the non-interacting case $U=U'=J=0$, the Green function,
${\bf G}_0=G_0 {\bf 1 }$, is given by
%%%%%%%%%%%%%%%%%%%%%%%%%%%%%%%%%%
\begin{eqnarray}
G_0^{-1}\left(z \right)&=&
z -\sum_k\frac{V_k^2}{z-\epsilon_k}.\label{eq:self1}
\end{eqnarray}
%%%%%%%%%%%%%%%%%%%%%%%%%%%%%%%%%%%%
When the interactions are introduced,
the full Green function ${\bf G}_{k}\left(z\right)$  is written as,
%%%%%%%%%%%%%%%%%%%%%%%%%
\begin{equation}
\left(
\begin{array}{cc}   		  
z-\Sigma_{\rm intra}(z)-\epsilon_k & -\Sigma_{\rm inter}(z)\\
-\Sigma_{\rm inter}(z) & z-\Sigma_{\rm intra}(z)-\epsilon_k
\end{array}
\right).
\end{equation}
%%%%%%%%%%%%%%%%%%%%%%%%%%%%%%%%%%%%%%%%%%
We have here used the fact that the self-energy ${\bf\Sigma}$ is independent 
of the momentum in $d\rightarrow\infty$ dimensions.
Then the diagonal element of the local Green function is given as,
%%%%%%%%%%%%%%%%%%%%%%%%%%%%%
\begin{eqnarray}
G_{\rm loc}\left(z\right)&=&\frac{1}{N}\sum_k G_{k}\left(z\right) 
\nonumber\\
&=&\int_{-\infty}^{\infty}
\frac{\left[z-\Sigma_{\rm intra}(z)-x\right]\rho\left(x\right)}
{\left(z-\Sigma_{\rm intra}(z)-x\right)^2-\Sigma_{\rm inter}(z)^2}dx,
\end{eqnarray}
%%%%%%%%%%%%%%%%%%%%%%%%%%%%%%%%%%%%%%%%
where $\rho(x)=\sum_{k} \delta\left(\epsilon_{k}-x\right)$ 
is the density of states in the non-interacting case.
The lattice structure is not so important to discuss the metal-insulator 
transition in $d=\infty$ dimensions.
We use here the semicircular density of states
$\rho(x)=\frac{2}{\pi D}\sqrt{1-(x/D)^2}$, 
which corresponds to an infinite-coordination Bethe lattice.
From the Dyson equation ${\bf G}_0^{-1}={\bf G}^{-1}+{\bf \Sigma}$, 
we then obtain the self-consistent 
equation as,
%%%%%%%%%%%%%%%%%%%%%%%%%%%%%%%%%%%%%%%%%%%%%%%%
\begin{eqnarray}
G_0^{-1}\left(z\right)&=&
z-\left(\frac{D}{2}\right)^2 G_{\rm loc}\left(z\right).\label{eq:self2}
\end{eqnarray}
%%%%%%%%%%%%%%%%%%%%%%%%%%%%%%%%%%%%%%%%%%%%%
We wish to note that the self-energy does not appear in these self-consistent 
equations, allowing us to simplify the iteration procedure.
Namely, by estimating the diagonal element of the local Green function for 
the Anderson model eq. (\ref{eq:And}) and by using
the self-consistent 
eqs. (\ref{eq:self1}) and (\ref{eq:self2}), we can discuss
Fermi-liquid properties in the system.			  
In the following, we take the band width $D$ as the unit of the
energy.

%%%%%%%%%%%%%%%%%%%%%%%%%%%%%%%%%%%%%%%%%%%%%%%%%%%%%%%%%%%%%%%%%%%%%%
\section{Metal-Insulator Transition}\label{sec3}
%%%%%%%%%%%%%%%%%%%%%%%%%%%%%%%%%%%%%%%%%%%%%%%%%%%%%%%%%%%%%%%%%%%%%%

To solve the self-consistent equations in DMFT,
we use the exact diagonalization method.
The diagonal element of the Green function for the Anderson  
Hamiltonian eq. (\ref{eq:And}) is given as,
%%%%%%%%%%%%%%%%%%%%%%%%%%%%%%%%%
\begin{eqnarray}
%G\left(z\right)&=&\sum_m \left[
%\frac{\left| \langle m|f_{1\uparrow}^\dag|0\rangle \right|^2}{z+E_0-E_m}+
%\frac{\left|\langle m|f_{1\uparrow}|0 \rangle \right|^2}{z-E_0+E_m}
%\right],
G\left(z\right)&=&\langle g|f_{1\uparrow}
\frac{1}{z+E_0-H}f_{1\uparrow}^\dag|g \rangle
\nonumber\\
&&+\langle g|f_{1\uparrow}^\dag
\frac{1}{z-E_0+H}f_{1\uparrow}|g \rangle\nonumber\\
&=&f\left(z+E_0, f_{1\uparrow}^\dag|g \rangle \right)\nonumber\\
&&-f\left(-z+E_0, f_{1\uparrow}|g \rangle \right).
\end{eqnarray}
%%%%%%%%%%%%%%%%%%%%%%%%%%%%%%%%%%%%%%%%
where $E_0$ is the ground state energy. 
The function $f\left(\omega, |\psi\rangle \right)$
is given by the following continued fraction\cite{Haydock},
%%%%%%%%%%%%%%%%%%%%%%%%%%%%%%%%%%%%%%%%
\begin{eqnarray}
f\left(\omega, |\psi\rangle
 \right)&\equiv& \langle\psi|\frac{1}{\omega-H}|\psi\rangle\nonumber\\
&=&\frac{\displaystyle \langle \psi|\psi\rangle}{\omega-a_1-
\frac{\displaystyle b_1^2}{\displaystyle \omega-a_2-\frac{
\displaystyle b_2^2}{\displaystyle \cdots}}},
\end{eqnarray}
%%%%%%%%%%%%%%%%%%%%%%%%%%%%%%%%%%%%%%%%%%
where $a_n$ ($b_n$) is the $n$th diagonal (subdiagonal) element of 
the Hamiltonian $H$ tridiagonalized by the Lanczos method 
with the initial vector $|\psi\rangle$.
To solve the self-consistent eqs. (\ref{eq:self1}) and (\ref{eq:self2})
in terms of  the obtained Green function $G$, 
we introduce the following cost function $\chi^2$ proposed by 
Caffarel and Krauth\cite{Caffarel} as 
%%%%%%%%%%%%%%%%%%%%%%%%%%%%%
\begin{eqnarray}
&&\frac{1}{n_{max}+1}\sum_{n=0}^{n_{max}}\left|
\left[G_0^{\rm old}\left(i\omega_n\right)\right]^{-1}-
\left[G_0^{\rm new}\left(i\omega_n\right)\right]^{-1}\right|^2,\label{error}
\end{eqnarray}
%%%%%%%%%%%%%%%%%%%%%%%%%%%%
where $n_{max}$ is chosen sufficiently large, and 
$\omega_n(=\frac{2n+1}{\tilde{\beta}}\pi)$ is the Matsubara frequency.
In terms of the conjugate gradient method, we determine the parameters 
$\epsilon_k$ and $V_k$ in $G_0^{\rm new}$ by minimizing 
 $\chi^2$ in eq. (\ref{error}).
To properly converge the iteration in DMFT, we introduce 
the "fictitious" inverse temperature $\tilde{\beta}$, which
is fixed  as $\tilde{\beta}=50$ in this paper. The number 
of site is set as $N=6$. We note that
a careful scaling analysis for the fictitious temperature and 
the number of sites is necessary only when  the system is near 
the metal-insulator transition point.
To discuss the stability of the Fermi liquid state, we define the 
wave-function renormalization 
factor in terms of the self-energy around  $\omega=0$, 
%%%%%%%%%%%%%%%%%%%%%%%%%%%%%%%%%%%%%%%%%%%%%
\begin{eqnarray}
Z&=&\left[1-\frac{\Delta {\rm Im} 
\Sigma\left(i\omega\right)}{\Delta\omega}\right]^{-1},
\end{eqnarray}
%%%%%%%%%%%%%%%%%%%%%%%%%%%%%%%%%%%%%%%%%%%%%%%%%%%%,
which corresponds to the weight of a quasiparticle excitation.

%%%%%%%%%%%%%%%%%%%%%%%%%%%%%%%%%%%%%%%%%%%%%%%%%%%%%%%%%%%%%%%%%%%%%%%%%%%
\subsection{Interplay of intra- and inter-band interactions}
%%%%%%%%%%%%%%%%%%%%%%%%%%%%%%%%%%%%%%%%%%%%%%%%%%%%%%%%%%%%%%%%%%%%%%%%%%%

Let us first discuss  the Fermi liquid properties
 of a metallic state in the absence 
of the Hund coupling $(J=0)$, and  clarify how the 
orbital-degeneracy affects a metal-insulator transition.
By making use of the exact diagonalization method, we iterate the
procedure mentioned above  to obtain the results within 
desired accuracy.

\begin{figure}[htb]
\begin{center}
\includegraphics[width=7cm]{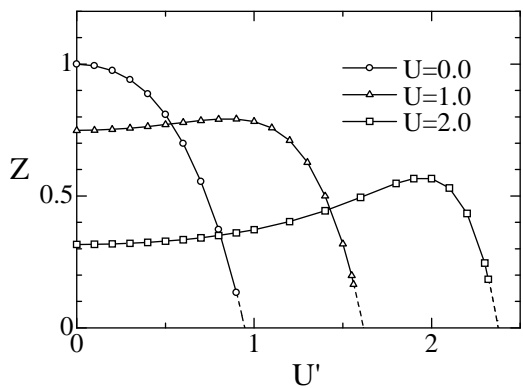}
\end{center}
\vskip -4mm
\caption{The quasi-particle weight $Z$ as a function of the inter-band Coulomb 
interaction $U'$, obtained by  the exact 
diagonalization $(N=6)$ within DMFT.}
\label{fig:Z1}
\end{figure}

We show the quasi-particle weight $Z$ as a function of the 
interband interaction 
$U'$ in Fig. \ref{fig:Z1}.
%%%%%%%%%%%%%%%%%%%%%%%%%%%%%%%%%%%%%%%%%%%%%%%%%%%%%%%%%%%%%%%%%%
%%%%%%%%%%%%%%%%%%%%%%%%%%%%%%%%%%%%%%%%%%%%%%%%%%%%%%%%%%%%%%%
In the case $U=0$, $Z$ decreases monotonically with 
increasing $U'$, and a metal-insulator transition occurs around
 $U'_c\sim 0.9$.  On the other hand,
 when $U\neq 0$, there appears nonmonotonic behavior
 in $Z$, namely, it once increases on the 
introduction of the interband Coulomb interaction,
 has the maximum value  in the vicinity of $U'\sim U$, 
and finally leads to a metal-insulator transition at a 
 critical value of $U'$. 
It is easily understood that the large value of  $U'$ 
suppresses hopping among sites, thereby causing a sharp
decrease in the renormalization factor $Z$.
However, it is somehow unexpected that the maximum structure 
appears around $U\sim U'$, and furthermore is more enhanced
for larger $U$ and $U'$.  We think that this 
is related to orbital fluctuations, which may
stabilize a normal metallic state around  $U \sim U'$.

\begin{figure}[htb]
\begin{center}
\includegraphics[width=7cm]{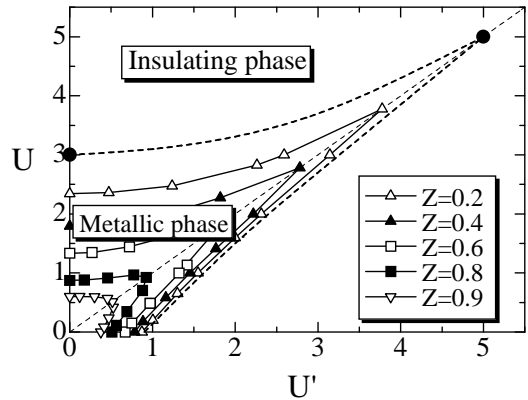}
\end{center}
\vskip -4mm
\caption{The contour plot of the quasi-particle weight $Z$ in the case $J=0$.
The bold-dashed line represents the phase boundary of the metal-insulator 
transition, which is deduced  by estimating 
the values of $U$ and $U'$ that give $Z=0$.
%%which is drawn as guide to eyes. 
The solid circles are the transition points obtained by the linearized DMFT.
}
\label{fig:phaseJ0}
\end{figure}

In order to observe the above behavior more clearly,
we show the contour plot of the quasi-particle weight in Fig. 
\ref{fig:phaseJ0}.
%%%%%%%%%%%%%%%%%%%%%%%%%%%%%%%%%%%%%%%%%%%%%%%%%%%%%%%%%%%%%%%%%%
%%%%%%%%%%%%%%%%%%%%%%%%%%%%%%%%%%%%%%%%%%%%%%%%%%%%%%%%%%%%%%%
When $U'=0$, the system is reduced to the single-band Hubbard model, 
where the intra-band Coulomb interaction $U$ induces 
the metal-insulator transition at $U_c$, which was estimated by 
the numerical renormalization group $(U_c=2.94)$,\cite{BullaNRG}
the exact diagonalization $(U_c=2.93)$,\cite{OnoED} 
the linearized DMFT $(U_c=3)$.\cite{Bulla}  These values 
agree with the results in Fig. \ref{fig:phaseJ0} well.

There are several remarkable features in this phase diagram.
We first notice that the value of $Z$ is not so sensitive to
$U'$ for a given $U$ ($>U'$), except for the 
region $U \sim U'$. In particular, the phase boundary indicated by 
the dashed line in the upper side of the figure
 is almost flat for the small $U$ region. 
 An important point to be noted is that when $U \sim U'$ the metallic state
is remarkably stable against a transition to the insulator,  and 
persists up to fairly large Coulomb interactions.  Moreover, 
it becomes immediately  unstable,
once the parameters are away from the line $U=U'$.  
This tendency becomes more conspicuous in the regime of  strong correlations.
These remarkable properties about the metallic phase
around the line $U=U'$ are closely related to orbital fluctuations,
as we will see momentarily in the following.

The second point is that there are two insulating 
phases corresponding to  the regions $U>U'$ and $U<U'$, which 
are separated by the metallic phase for small $U, U'$, but 
are continuously connected 
to each other via  the region of large $U, U'$.
For example, let us observe phase transitions with  $U=4.0$ being fixed.
When $U'=0.0$ with $U=4.0$, the system is 
in the insulating phase caused by the intra-band interaction $U$. 
Introducing the interband coupling $U'$, a phase transition occurs to the 
metallic phase, and  further increase in $U'$  induces the second
transition to another insulating phase, which is 
dominated by the inter-band interaction $U'$. These two insulating
phases should show different properties, though they can be 
adiabatically connected to each other in principle.

%%%%%%%%%%%%%%%%%%%%%%%%%%%%%%%%%%%%%%%%%%%%%%%%%%%%%%%%%%%%%%%%%%%
\subsection{Critical properties around the phase boundary}
%%%%%%%%%%%%%%%%%%%%%%%%%%%%%%%%%%%%%%%%%%%%%%%%%%%%%%%%%%%%%%%%%%%

To clarify the above characteristic 
properties around the metal-insulator transition,
 we exploit a linearized version of DMFT,\cite{Bulla}  
 which provides us with a transparent view of the 
 phase transitions. In this scheme,
we introduce  a specific Anderson impurity model
connected to only {\it one host site}, which corresponds to the model 
eq. (\ref{eq:And}) with $k=1$.  This extreme simplification 
even provides  sensible results as far as low-energy 
properties near the transition point are concerned.
Namely, for low-energy  excitations around the Fermi surface,
the Green function for $f$-electrons may be approximated by a single 
pole as,
%%%%%%%%%%%%%%%%%%%%%%%%%
\begin{eqnarray} 
G_{\rm loc}(z)&\sim& \frac{w}{z},
\end{eqnarray} 
%%%%%%%%%%%%%%%%%%%%%%%%%%%%%%%%%%%
where the residue $w$ corresponds to the quasi-particle weight.
Recall that for $w=0$  the ground state belongs to an
insulating phase, while for $w\neq0$ to a metallic phase.
We can thus determine the metal-insulator transition
point from the condition $w=0$, as has been done before.
%%%%%%%%%%%%%%%%%%%%%%%%%%%%%%%%%%%%%%%%%%%%%%%%%%%%%%%%%%%%%%%%%%
\begin{figure}[htb]
\begin{center}
\includegraphics[width=7cm]{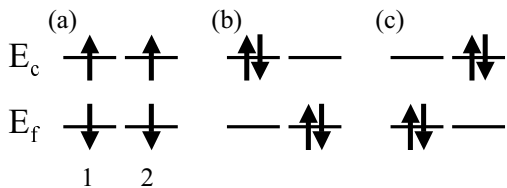}
\end{center}
\vskip -4mm
\caption{Three distinct singlet states in the 
 Anderson model, which is hybridized with only
one effective host site.  The numbers 1, 2 
specify the orbital indices.
}
\label{fig:ponch}
\end{figure}
%%%%%%%%%%%%%%%%%%%%%%%%%%%%%%%%%%%%%%%%%%%%%%%%%%%%%%%%%%%%%%%
By combining the self-consistent eqs. (\ref{eq:self1}) 
and (\ref{eq:self2}), we iterate the linearized DMFT.
The self-consistent condition imposed on the hybridization
now reads
%%%%%%%%%%%%%%%%%%%%%%%%%%%%%%%%%%%%%%%%%%%%%%%%%%
\begin{eqnarray}
V &=&\frac{\sqrt{w}}{2}.\label{eq:iter}
\end{eqnarray}
%%%%%%%%%%%%%%%%%%%%%%%%%%%%%%%%%%%%%%%%%%%%%%
For the single-band Hubbard model ($U'=0$), 
the critical value $U_c=3$ was already estimated by 
the linearized DMFT.\cite{Bulla}

In this simplified model, the metallic ground state is in 
 the spin singlet state drawn in Fig. \ref{fig:ponch} (a) 
schematically.
Although the introduction of the interband Coulomb interaction $U'$
lowers  the states (b) and (c)  energetically,  
 these states do not affect the ground-state properties
 up to a certain critical value of $U'$.
In fact, the relevant excitation gap is independent of $U'$
in this approximation, and 
the residue of the Green function around $V=0$ is given as,
%%%%%%%%%%%%%%%%%%%%%%%%%%%%%%%%%%%%%%%%%%%%%
\begin{eqnarray}
w&=&36\left(\frac{V}{U}\right)^2,
\end{eqnarray}
%%%%%%%%%%%%%%%%%%%%%%%%%%%%%%%%%%%%%%%%%%%%%
via a straightforward calculation.
Then the self-consistent eq. (\ref{eq:iter}) updates 
 the hybridization via $(3/U) V \rightarrow V$ 
 in each iteration process. Therefore, 
when $U>U_c(=3)$, the effective hybridization $V$ 
 vanishes by the iteration, thereby
stabilizing the insulating phase in this region.
On the other hand, the hybridization is 
relevant for $U<U_c$ and thus
the metallic phase is realized.  Note that
the critical value $U_c=3$ obtained here is the same as that for 
the single-band Hubbard model.\cite{Bulla}
We thus come to the conclusion  that 
 the inter-band Coulomb interaction has little effect 
on the  metal-insulator transition, which is essentially 
described by the mechanism for 
the single-band Hubbard model,  as far as the case of $U> U'$
is concerned.
This statement is indeed consistent with the numerical results
 shown in Fig. \ref{fig:phaseJ0}, namely,  the 
 phase boundary is rather insensitive to the change in  $U'$
 for small $U'$.

On the other hand, in the special case $U=U'$, the nature of the 
transition is totally changed in this approximation. In this case, 
the ground state is formed by three degenerate states shown in Fig. 
 \ref{fig:ponch}, and not only spin but also 
  orbital degrees of freedom play an important role.  
In fact, quantum fluctuations among these states decrease 
the effective mass, making  the metallic phase more stable.
The critical value $U_c=5$ estimated by the present scheme
agrees with the  numerical one in Fig. \ref{fig:phaseJ0}.
 Also this is consistent with the conclusion of quantum Monte Carlo 
simulations,\cite{Han} which claims that 
the ground state in the single (double) 
orbital Hubbard model belongs to the insulating (metallic) 
phase in the case  $U=U'=4$.  

The present analysis also gives a clear picture for 
 two types of the metal-insulator
 transitions: the insulating phase in the upper region of
  in Fig. \ref{fig:phaseJ0} 
 corresponds to the state of Fig. \ref{fig:ponch}(a), where the
 intra-atomic repulsion gives rise to the insulating phase, 
 whereas the insulator in the lower region, corresponding to
 Fig. \ref{fig:ponch}(b) and (c), 
  is  stabilized by the inter-band interaction. Around the critical point 
 on the $U=U'$ line, the metallic state sandwiched by these
 insulating phases is  realized by the competition among almost degenerate
  singlet states  (a), (b) and (c). In other words, the 
 metallic state is stabilized by orbital fluctuations.

%%%%%%%%%%%%%%%%%%%%%%%%%%%%%%%%%%%%%%
\section{Effects of the Hund Coupling}\label{sec4}
%%%%%%%%%%%%%%%%%%%%%%%%%%%%%%%%%%%%

%%%%%%%%%%%%%%%%%%%%%%%%%%%%%%%
\subsection{Isotropic case}
%%%%%%%%%%%%%%%%%%%%%%%%%%%%%%%%

\begin{figure}[htb]
\begin{center}
\includegraphics[width=7cm]{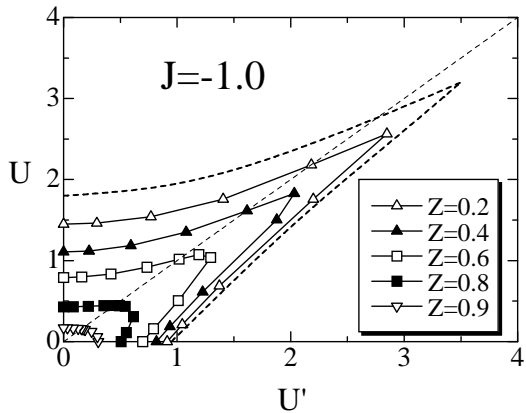}
\end{center}
\vskip -4mm
\caption{The contour plot of the quasi-particle weight $Z$ in 
the system with 
the ferromagnetic Hund coupling between the orbitals $J=-1.0$.
}
\label{fig:phase10}
\end{figure}

\begin{figure}[htb]
\begin{center}
\includegraphics[width=7cm]{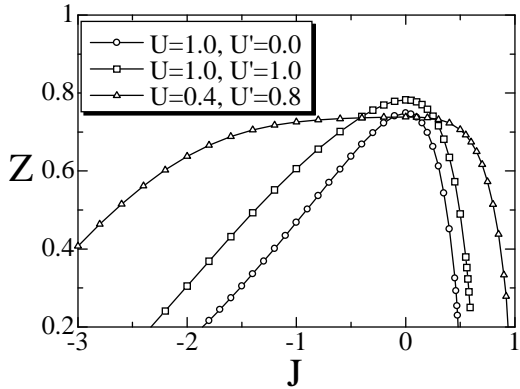}
\end{center}
\vskip -4mm
\caption{The quasi-particle weight $Z$ as a function of $J$.
}
\label{fig:J}
\end{figure}

We now discuss the effects of the exchange coupling 
$J$ between orbitals,
which plays an important role in real materials such as  manganites 
and ruthenates.
By performing similar calculations, we obtain the contour plot of the 
quasi-particle weight $Z$ as shown in Fig. \ref{fig:phase10}. 
We also plot the values of $Z$ as a function of $J$
in Fig. \ref{fig:J}.
%%%%%%%%%%%%%%%%%%%%%%%%%%%%%%%%%%%%%%%%%%%%%%%%%%%%%%%%%%%%%%%%%%
%%%%%%%%%%%%%%%%%%%%%%%%%%%%%%%%%%%%%%%%%%%%%%%%%%%%%%%%%%%%%%%
%%%%%%%%%%%%%%%%%%%%%%%%%%%%%%%%%%%%%%%%%%%%%%%%%%%%%%%%%%%%%%%
%%%%%%%%%%%%%%%%%%%%%%%%%%%%%%%%%%%%%%%%%%%%%%%%%%%%%%%%%%%%%%%
The Hund coupling $J$ ($<0$) enhances spin correlations
between the orbitals, thus renormalizing  electrons near 
the Fermi surface. It is indeed seen
 in both figures that for $U>U'$ 
 the quasi-particle 
weight is decreased with the increase of $|J|$.
On the other hand, the effect of the Hund coupling is less conspicuous 
 in the region $U<U'$.
In the latter region,  the inter-band Coulomb interaction 
dominates electron correlations in the ground state when $J=0$,
which is schematically drawn as Fig. \ref{fig:ponch} (b) or (c).
For these configurations the Hund coupling is irrelevant,
which explains that the effect of $J$ is 
less important for $U<U'$.
%%%%%%%%%%%%%%%%%%%%%%%%%%%%%%%%%%%%%%%%%%%%%%%%%%%%%%%%%%%%%
\begin{figure}[htb]
\begin{center}
\includegraphics[width=7cm]{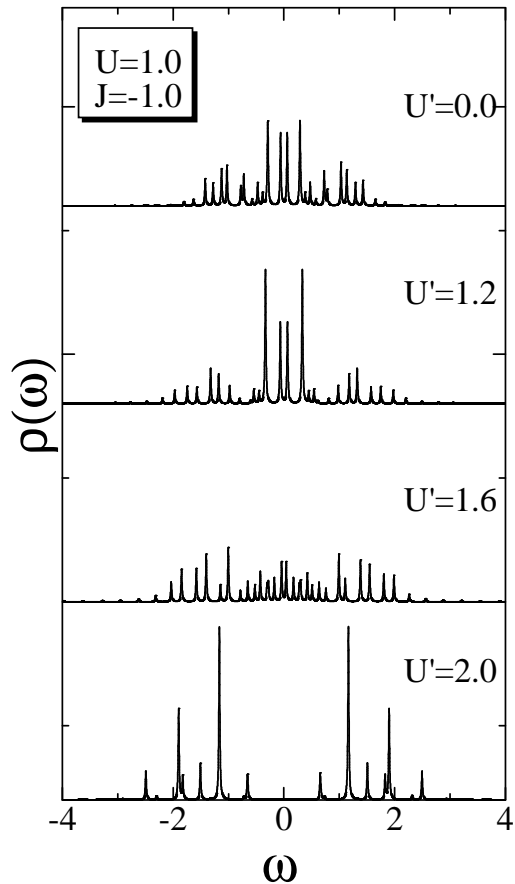}
\end{center}
\vskip -4mm
\caption{The density of states for $U'=0.0, 1.2, 1.6$ and $2.0$ obtained
by the exact diagonalization with $\delta=0.01$
}
\label{fig:rho}
\end{figure}
%%%%%%%%%%%%%%%%%%%%%%%%%%%%%%%%%%%%%%%%%%%%%%%%%%%%%%%%%%%%%%%

To make the metal-insulator transition clear, 
we also calculate the density of states 
by means of the exact diagonalization. The results 
are shown in Fig. \ref{fig:rho}.
When $U=1.0, J=-1.0$ and $U'=0.0$, the system belongs to the metallic phase, 
as shown in Fig. \ref{fig:phase10}.
Introducing the interband coupling $U'$, fluctuations between the orbitals are
enhanced, which assist to stabilize the quasi-particle states up to $U'\sim1.2$.
Further increase in $U'$  favors the doubly occupied state in
the same orbital. Therefore, the electron states around the Fermi surface 
are pushed away to higher energy regions, as seen in the case of $U'=1.6$,  
and finally a quantum phase transition occurs from the metallic phase 
to the insulating phase.
In the insulating phase for  $U'=2.0$, we can clearly see 
the charge gap in the density of states in Fig. \ref{fig:rho}.

%%%%%%%%%%%%%%%%%%%%%%%%%%%%%%%%%%%%%%%%%%%%%%%%%%%%%%%%%%%%%%%%%%
\begin{figure}[htb]
\begin{center}
\includegraphics[width=7cm]{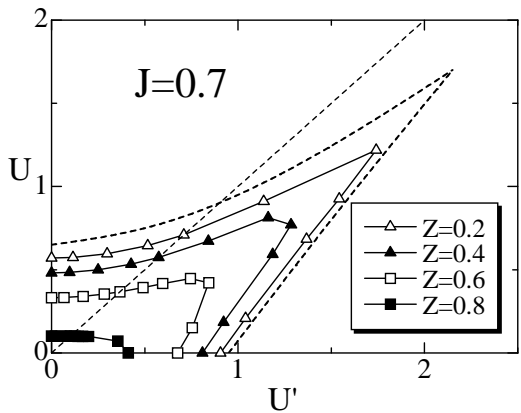}
\end{center}
\vskip -4mm
\caption{The contour plot of the quasi-particle weight $Z$ in the system with 
the antiferromagnetic exchange coupling between the orbitals $J=0.7$.
}
\label{fig:phase7}
\end{figure}
%%%%%%%%%%%%%%%%%%%%%%%%%%%%%%%%%%%%%%%%%%%%%%%%%%%%%%%%%%%%%%%

We also investigate the case of the antiferromagnetic
 exchange coupling $(J>0)$.
The contour plot in the case $J=0.7$ is shown in Fig. \ref{fig:phase7}.
By comparing with the ferromagnetic case, it is seen that
electrons are renormalized strongly, as shown in Fig. \ref{fig:J}.
This difference is explained by counting the number 
of effective degrees of freedom at each site in the presence of the 
exchange coupling. In the ferromagnetic case, 
the system has a tendency to form the  $S=1$ state at each site,
while in the antiferromagnetic case it favors 
the singlet state made of two orbitals.  The resulting
 effective degrees of freedom give rise to
the difference in the stability of  the Fermi liquid state:
the metallic state with the ferromagnetic coupling is more
 stable than  the antiferromagnetic case.
 Note that the situation is similar to the stability of 
 the metallic state due to orbital 
 fluctuations around $U \sim U'$ discussed in the previous section.

%%%%%%%%%%%%%%%%%%%%%%%%%%%%%%%%%%%%%%%%%%%%%%%%%%%%%%%%%%%%%%%%%%
\subsection{Anisotropic case }
%%%%%%%%%%%%%%%%%%%%%%%%%%%%%%%%%%%%%%%%%%%%%%%%%%%%%%%%%%%%%%%%%%

\begin{figure}[htb]
\begin{center}
\includegraphics[width=7cm]{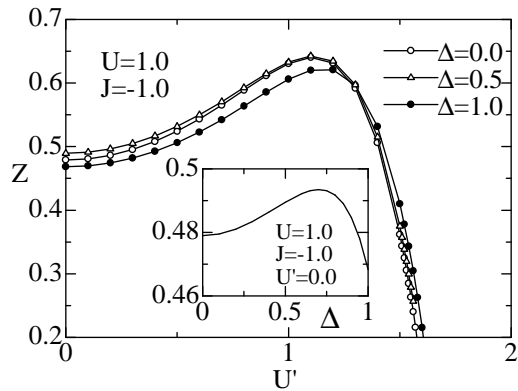}
\end{center}
\vskip -4mm
\caption{The quasi-particle weight $Z$ for the case with
anisotropic exchange coupling: $U=1.0$ and $J=-1.0$. 
Inset shows the quasi-particle weight 
as a function of the anisotropy parameter $\Delta$.
}
\label{fig:xxz}
\end{figure}

\begin{figure}[htb]
\begin{center}
\includegraphics[width=7cm]{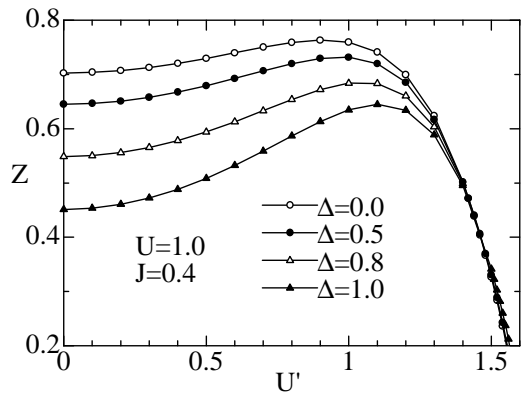}
\end{center}
\vskip -4mm
\caption{The quasi-particle weight $Z$ 
as a function of $U'$ in the case $U=1.0$.
}
\label{fig:xxz4}
\end{figure}

We finally discuss the effects of anisotropy in the Hund coupling.
For this purpose, we introduce the anisotropy 
parameter $\Delta$ as, 
%%%%%%%%%%%%%%%%%%%%%%%%%%%%%%%%%%%%%%%
\begin{eqnarray}
\left({\bf S}_1\cdot{\bf S}_2\right)_\Delta&=&
\Delta\left(S_1^x S_2^x+S_1^y S_2^y\right)+S_1^z S_2^z, 
\end{eqnarray} 
%%%%%%%%%%%%%%%%%%%%%%%%%%%%%%%%%%%%%%%%
where $\Delta=0$ corresponds to the Ising-type anisotropy,
while  $\Delta=1$ the isotropic case. 
In Fig. \ref{fig:xxz}, we show the quasi-particle weight $Z$
for  $U=1.0$  and $J=-1.0$.
%%%%%%%%%%%%%%%%%%%%%%%%%%%%%%%%%%%%%%%%%%%%%%%%%%%%%%%%%%%%%%%%%%
%%%%%%%%%%%%%%%%%%%%%%%%%%%%%%%%%%%%%%%%%%%%%%%%%%%%%%%%%%%%%%%
In the ferromagnetic case,  the quasi-particle weight 
$Z$ is little affected by anisotropy although 
somewhat nonmonotonic dependence is found as a function of  $\Delta$ 
as shown in the inset.
On the contrary, the anisotropy has a noticeable effect for 
 the antiferromagnetic case with small $U'$, as seen from 
 Fig. \ref{fig:xxz4}, where
 the anisotropy decreases the weight $w$
monotonically with the increase of $\Delta$.
 This difference arises 
from the fact that the $x-y$ components of the exchange coupling 
bring about quantum fluctuations more prominently in the
antiferromagnetic case than in the ferromagnetic case, 
giving rise to the sizable effect on the antiferromagnetic 
case in the small $U'$ region.

%%%%%%%%%%%%%%%%%%%%%%%%%%%%%%%%%%%%%%%%%%%%%%%%%%%%%%%%%%%%%%%%%%
%%%%%%%%%%%%%%%%%%%%%%%%%%%%%%%%%%%%%%%%%%%%%%%%%%%%%%%%%%%%%%%

%%%%%%%%%%%%%%%%%%%%%%%%%%%%%%%%%%%%
\section{Summary}\label{Summary}
%%%%%%%%%%%%%%%%%%%%%%%%%%%%%%%%%%%%

We have investigated the stability of a metallic phase 
in the two-orbital
Hubbard model at half-filling by means of dynamical mean-field theory
with particular emphasis on  the role played by 
orbital degrees of freedom. By combining
 the exact diagonalization  method with DMFT, we have 
discussed how 
the interplay among the intra-band and inter-band Coulomb interaction
together with the Hund coupling affects the metal-insulator transition. 
In particular, it has been found that the metallic state is 
remarkably stable, even up to $U\sim  5.0$
if  the intra and inter-band Coulomb interactions 
are nearly equal. Also, slight deviation from this condition 
immediately drives the  system to the insulating phase. 
We have pointed out that orbital fluctuations play 
a particular role to realize the metallic state around
 $U \sim U' \sim 5.0$.
The effect of the exchange coupling has been also discussed.  			
It has been clarified that the effective degrees of freedom
 at each site play an important role again 
 in stabilizing the metallic state. 
 
We have focused on nonmagnetic phases of the model in this paper.
Possible instabilities toward other ordered states such as the magnetic order, 
the superconductivity, etc., are to be studied.
In particular, the ferromagnetic instability is interesting 
in connection with ferromagnetism realized in manganite compounds.
These problems are now under consideration. 

%%%%%%%%%%%%%%%%%%%%%%%%%%%%%%%%%%%%%%%%%%%%%%%%%%%%%%%%%%%%%%%
\section{Acknowledgments}
We would like to thank M. Suminokura for useful discussions.
This work was partly supported by a Grant-in-Aid from the Ministry 
of Education, Science, Sports and Culture of Japan. 
A part of computations was done at the Supercomputer Center at the 
Institute for Solid State Physics, University of Tokyo
and Yukawa Institute Computer Facility.

%%%%%%%%%%%%%%%%%%%%%%%%%%%%%%%%%%%%%%%%%%%%%%%%%%%%%%%%%%%%%%%%%%%%%
%                        REFERENCES                                 %
%%%%%%%%%%%%%%%%%%%%%%%%%%%%%%%%%%%%%%%%%%%%%%%%%%%%%%%%%%%%%%%%%%%%%
%

%%%


\begin{thebibliography}{99}

\bibitem{Tokura}
Y. Tokura, A. Urushibara, Y. Moritomo, T. Arima, A. Asamitsu, G. Kido and 
N. Furukawa, J. Phys. Soc. Jpn. {\bf 63}, 3931 (1994).

\bibitem{Maeno}
Y. Maeno, H. Hashimoto, K. Toshida, S. Nishizaki, T. Fujita, J. G. Bednorz, 
F. Lichtenberg, Nature, {\bf 372}, 532 (1994).
			   
\bibitem{Tokura2}
M. Imada, A. Fujimori and Y. Tokura,
Rev. Mod. Phys. {\bf 70}, 1039 (1998);
Y. Tokura and N. Nagaosa, Science {\bf 288}, 462 (2000).

\bibitem{RiceSigrist}
T. M. Rice and M. Sigrist, J. Phys. Condens. Matt. {\bf 7}, L643 (1995).

\bibitem{Kondo}
S. Kondo, D. C. Johnston, C. A. Swenson, F. Borsa, A. V. Mahajan, L. L. Miller,
T. Gu, A. I. Goldman, M. B. Maple, D. A. Gajewski, E. J. Freeman, N. R. Dilley,
R. P. Dickey, J. Merrin, K. Kojima, G. M. Luke, Y. J. Uemura, O. Chmaissem 
and J. D. Jorgensen, Phys. Rev. Lett. {\bf 78}, 3729 (1997).

\bibitem{Kaps}
H. Kaps, N. B\"uttgen, W. Trinkl, A. Loidl, M. Klemm and S. Horn,
J. Phys. Condense. Matter, {\bf 13}, 8497 (2001).

\bibitem{Isoda}
M. Isoda and S. Mori, J. Phys. Soc. Jpn. {\bf 69}, 1509 (2000).

\bibitem{Fujimoto}
S. Fujimoto, Phys. Rev. B {\bf 64}, 085102 (2001).

\bibitem{Tsunetsugu}
H. Tsunetsugu, preprint.

\bibitem{Yamashita}
Y. Yamashita and K. Ueda, preprint.

%%%%%%%%%%%%%%%%% orbital 

\bibitem{Hasegawa}
H. Hasegawa, J. Phys. Soc. Jpn. {\bf 66}, 1391 (1997).

\bibitem{Ishihara}
S. Ishihara, M. Yamanaka and N. Nagaosa, Phys. Rev. B {\bf 56}, 686 (1997);
R. Maezono, S. Ishihara and N. Nagaosa, Phys. Rev. B {\bf 58}, 11583 (1998).

\bibitem{Fresard}
R. Fr\'esard and G. Kotliar, Phys. Rev. B {\bf 56},12909 (1997).

\bibitem{Klenjnberg}
A. Klejnberg and J. Spalek, Phys. Rev. B {\bf 57}, 12041 (1998).

\bibitem{Bunemann}
J. B\"unemann, W. Weber and F. Gebhard, Phys. Rev. B {\bf 57}, 6896 (1998).

\bibitem{Motome}
Y. Motome and M. Imada, J. Phys. Soc. Jpn. {\bf 67}, 3199 (1998).

\bibitem{Khaliullin}
G. Khaliullin and S. Maekawa, Phys. Rev. Lett. {\bf 85}, 3950 (2000).   	

\bibitem{Takimoto}
T. Takimoto, Phys. Rev. B {\bf 62}, R14641 (2000).
%%%%%%%%%%%%%%%%%%%%%%%%%%%  

% DMFT 
\bibitem{Momoi}
T. Momoi and K. Kubo, Phys. Rev. B {\bf 58}, R567 (1998).

\bibitem{Rozenberg}
M. J. Rozenberg, Phys. Rev. B {\bf 55}, R4855 (1997).

\bibitem{Held}
K. Held and D. Vollhardt, Eur. Phys. J. B {\bf 5}, 473 (1998).

\bibitem{Han}
J. E. Han, M. Jarrell and D. L. Cox, Phys. Rev. B {\bf 58}, R4199 (1998).

\bibitem{QMC}
V. S. Oudovenko and G. Kotliar, Phys. Rev. B {\bf 65}, 075102 (2002).

	 
\bibitem{Kotliar}
G. Kotliar and H. Kajueter, Phys. Rev. B {\bf 54}, R14221 (1996).
%%%%%%%%%%%%%%%%%%%%%%%%%%
\bibitem{Metzner}
W. Metzner and D. Vollhardt, Phys. Rev. Lett. {\bf 62}, 324 (1989).

\bibitem{Muller}
E. M\"uller-Hartmann, Z. Phys. B {\bf 74}, 507 (1989).

\bibitem{Georges}
A. Georges, G. Kotliar, W. Krauth and M. J. Rozenberg,
Rev. Mod. Phys. {\bf 68}, 13 (1996).

\bibitem{Pruschke}
T. Pruschke, M. Jarrell, and J. K. Freericks, 
Adv. Phys. {\bf 42}, 187 (1995).

\bibitem{Caffarel}
M. Caffarel and W. Krauth, Phys. Rev. Lett. {\bf 72}, 1545 (1994).  	

\bibitem{single1}
Th. Pruschke, D. L. Cox and M. Jarrell:  Phys. Rev. B 47 (1993) 3553.

\bibitem{Sakai}
O. Sakai and Y. Kuramoto, Solid State Comm. {\bf 89}, 307 (1994).

\bibitem{single2}
R. Chitra and G. Kotliar, Phys. Rev. Lett. {\bf 83}, 2386 (1999).

\bibitem{single3}
J. Joo and V. Oudovenko, Phys. Rev. B {\bf 64}, 193102 (2001).

\bibitem{single4}
M. S. Laad, L. Craco and E. M\"uller-Hartmann,
Phys. Rev. B {\bf 64}, 195114 (2001). 

\bibitem{BullaNRG}
R. Bulla, Phys. Rev. Lett. {\bf 83}, 136 (1999).

\bibitem{2band1}
A. Georges, G. Kotliar and W. Krauth, Z. Phys. B {\bf 92}, 313 (1993).

\bibitem{2band2}
Th. Maier, M. B. Z\"olfl, Th. Pruschke and J. Keller, 
Eur. Phys. J. B {\bf 7}, 377 (1999).

\bibitem{2band3}
Y. Imai and N. Kawakami, J. Phys. Soc. Jpn. {\bf 70}, 2365 (2001).     

\bibitem{OnoED}
Y. Ono, R. Bulla and A. C. Hewson, Eur. Phys. J. B {\bf 19}, 375 (2001);
Y. Ohashi and Y. Ono, J. Phys. Soc. Jpn. {\bf 70}, 2989 (2001).

\bibitem{PAM}
M. Jarrell, H. Akhlaghpour and T. Pruschke,
Phys. Rev. Lett. {\bf 70}, 1670 (1993). 

\bibitem{Mutou}
T. Mutou and D. Hirashima, J. Phys. Soc. Jpn. {\bf 63}, 4475 (1994);

\bibitem{Saso}
T. Saso and M. Itoh, Phys. Rev. B {\bf 53}, 6877 (1996);  
%T. Muto, Phys. Rev. B {\bf 64}, 165103 (2001); {\bf 64}, 245102 (2001).


  



\bibitem{Bulla}
R. Bulla and M. Potthoff, Eur. Phys. J. B {\bf 13}, 257 (2000).

\bibitem{Haydock}
R. Haydock, V. Heine and M. J. Kelly, J. Phys. C {\bf 8}, 2591 (1975).  	


\end{thebibliography}
\end{document}